\documentclass[]{aa}
\usepackage{graphicx}
\usepackage{natbib}
\usepackage{amssymb}
\usepackage{hyperref}
\usepackage{txfonts}
\usepackage{rotating,bm}
\usepackage{amsmath}
\usepackage{natbib}

\begin{document}
\title{A method to deconvolve stellar rotational velocities II}
\subtitle{The probability distribution function via Tikhonov regularization}
\titlerunning{Tikhonov's rotational velocities distribution}
\author{
 Alejandra Christen\inst{1}
\and
Pedro Escarate \inst{2,3}
\and
Michel Cur\'{e}  \inst{4}
\and
Diego F. Rial \inst{5}
\and
Julia Cassetti \inst{6}
}
\institute{
Instituto de Estad\'{i}stica, Pontificia Universidad Cat\'{o}lica de Valpara\'{i}so, Chile,\\ \email{alejandra.christen@pucv.cl}
\and
Centro Avanzado de Ingenier\'ia El\'ectrica y Electr\'onica, Universidad T\'ecnica Federico Santa Mar\'ia, Chile, \email{pedro.escarate@usm.cl}
\and
Large Binocular Telescope Observatory, Steward Observatory, Tucson, AZ, USA
\and
Instituto de F\'{i}sica y Astronom\'{i}a, Universidad de Valpara\'{i}so, Chile \email{michel.cure@uv.cl}
\and
Departamento de Matem\'{a}ticas, Facultad de Ciencias Exactas y Naturales, Universidad de Buenos Aires, Argentina \email{drial@mate.uba.ar}
\and
Universidad Nacional de General Sarmiento, Buenos Aires, Argentina, \\
\email{jcassett@ungs.edu.ar}
}
\date{Received ; Accepted }
\abstract{} {Knowing the distribution of stellar rotational velocities is essential for the understanding
stellar evolution. Because we measure the projected rotational speed $v \sin i$, we need to solve 
an ill--posed problem given by a Fredholm integral of the first kind to recover the 'true' rotational velocity distribution.} {After discretization of the Fredholm integral, we apply the Tikhonov regularization method to
obtain directly the probability distribution function for stellar rotational velocities. We propose a simple and straightforward procedure to determine the Tikhonov parameter.  We applied Monte Carlo simulations to prove that Tikhonov method is a consistent estimator and asymptotically unbiased.} {This 
method is applied to a sample of cluster stars. We obtain confidences intervals using bootsrap method. Our results are in good agreement with the one obtained using the Lucy method, in recovering the probability density distribution of rotational velocities. Furthermore, Lucy estimation lies inside our confidence interval. } {Tikhonov regularization is a very robust method that deconvolve the rotational velocity probability density function from a sample of $v \sin i$ data straightforward without needing any convergence criteria.}
\keywords{
methods: analytical -- methods: data analysis -- methods: numerical -- methods: statistical -- 
stars: fundamental--parameters -- stars: rotation
}
\maketitle

\section{Introduction\label{sec:intro}}

The understanding about how stars rotate is essential to describe and modelling many aspect of stellar evolution. 
From spectroscopy observations we can only get the projected velocity, $v \sin i$, where $i$ is the inclination angle with respect to the line of sight. Furthermore, in order to deconvolve (disentangle or unfold) the rotational velocity distribution function, an assumption on the distribution of
rotational axes is required. The standard choice is that the distribution of stellar axes is uniformly (randomly) 
distributed over the sphere. Using this assumption Chandrasekhar \&  M\"unch (1950) studied the integral equation that describe the distribution of 'true'  ($v$) and  apparent ($v \sin i$) rotational velocities, deriving a formal solution, which is proportional to a derivative of an Abel's Integral. 
Chandrasekhar \& M\"unch (1950) method is not usually applied, because the differentiation of the formal 
solution can lead to misleading results due to intrinsic numerical problems associated to the derivative of the Abel's integral. 

Cur\'e et al. (2014) extended the work of Chandrasekhar \& M\"unch (1950), integrating the formal solution and obtained 
the cumulative distribution function (CDF) for the rotational velocities. This CDF is attained in one step demonstrating the robustness to this method.

While the CDF identifies the distribution of the speed of rotation it is sometimes useful to have the probability density function (PDF) for easy handling and to appreciate directly certain properties of the distribution (e.g., the maximum, its  symmetry, variability, etc). It is known also that the observed values of the projected rotational velocities are provided with measurement error. The goal of this work is to propose a methodology that provides straightforward the PDF, taking into account the measurement errors and avoiding numerical problems arising from the derivative of the CDF. 
Regularization methods are a technique widely used to deconvolve inverse problems. Image processing, geophysics and  machine learning are some of the areas where they are usually applied (Bouhamidi $2007$, Deng et al. $2013$, Fomel $2007$). Among the regularization methods we find: Truncated Singular Value Decomposition (TSVD), Selective Singular Value Decomposition (SSVD) and Tikhonov Regularization Method (Hansen 2010).

In this article we obtain the estimated probability distribution function directly from the Fredholm integral by means of the Tikhonov regularization method.

After its introduction by Tikhonov (1943) to solve integral equation problems, this method (known as Ridge Regression in statistics) has been developed and extensively used since then (see, e.g., Tikhonov 1963, Tikhonov and Arsenin 1977, Tikhonov et al. 1995, Eggermont 1993, Hansen 2010). It allows an increase in the numerical stability and dealing with errors of measurement.

This article is structured as follows: In section 2  we briefly present the mathematical description of the method and describe a procedure to calculate the Tikhonov factor. In section 3, we perform Monte Carlo simulations to show the robustness of this method. In section 4,  a real sample of cluster stars are deconvolved by Tikhonov regularization,  confidence intervals are calculated using bootstrap method and a comparison between our PDF results with the one obtained with the Lucy (1974) method and CDF results from the work of Cur\'e et al. (2014) are performed. Last section presents our conclusions and future work.

\section{Tikhonov Regularization Method\label{sec:method}}

Many inverse problems in physics and astronomy are given in terms of the Fredholm integral of the first kind (Lucy 1994, Hansen 2010), namely:
\begin{equation}
\label{eq1}
f_{Y}(y)=\int p(y\,|\,x)\,f_{X}(x)dx,
\end{equation}
here $f_{Y}$ is a function accessible to observation and $f_{X}$ is the function of interest. The kernel $p(y\,|\,x)$ of this integral is related to the remoteness of the measurement process, in this case, the projection of the distribution of stellar axes.

Chandrasekhar \& M\"unch (1950) were the first in considering the integral equation governing the distribution of 'true' an the apparent (projected)  rotational velocities of stars, $y=x\sin i$, where $x=v$ is the rotational speed and $i$ is the inclination angle with respect to the line of sight.
Assuming an  uniform distribution of stellar axes over the sphere (see Cur\'e et al. 2014 for details), this integral equation (Eq. \ref{eq1}) reads as  follows:

\begin{equation}
\label{eq-problema}
f_{Y}(y)=\int_{y}^{\infty}\dfrac{y}{x\sqrt{x^{2}-y^{2}}}\,f_{X}(x)dx.
\end{equation}

Expressing Eq. (\ref{eq-problema}) in matrix form (by a quadrature discretization of the problem), we get:

\begin{equation} \label{eq-ill}
Y\,=\, A \, X
\end{equation}

Now, $A$ is a matrix representing the kernel $p(y|x)$, $Y$ is a vector representing the density of projected rotational velocities $f_Y(y)$ and $X$ is the unknown vector representing the density of 'true' rotational velocities $f_X(x)$.\\

Since the observed data are measured with error, last equation is an example of a discrete ill--posed problem, i.e., small errors in the measured data can produce large 
variations in the recovered function which make the solution unstable (Ivanov et al. 2002 and references therein).
Nevertheless, in the decades after the work of Chandrasekhar \&  M\"unch (1950), much mathematical work on this kind of problems has been developed. Among them, one of the most common methods is the Tikhonov regularisation method 
(Tikhonov \& Arsenin 1977, Tikhonov et al. 1995, Hansen 2010). 

The standard method to solve Eq. (\ref{eq-ill}) is to apply ordinary least squares (OLS), i.e., $\min\{||A\,X-Y||^2\}$, where $||\cdot||$ represents the euclidean norm, but for ill--posed problems this method fails in the sense that can produce unstable estimators. In order to avoid this problem Tikhonov regularization method 
imposes a regularization term to be included in the minimization process, namely: 
\begin{equation}\label{eq-regterm}
\min\{||A\,X-Y||^2\} \rightarrow \min\{||A\,X-Y||^2 + \lambda^2 \,|| L \,(X-X_0) ||^2\},
\end{equation}
where $\lambda$ is the Tikhonov factor. The standard definition for the $L$ matrix is $L=I$, where $I$ is the identity matrix and $X_0$ is an initial estimation, setting $X_0 = 0$, when there is no previous information. There exist different quantitative approaches to obtain Tikhonov factor, e.g.,  Generalized Cross-Validation (GCV), L-curve Method, Discrepancy Principle, Restricted Maximum Likelihood. More details of these are explained in, e.g., Press et al. (2007), Hansen (2010), Tikhonov \& Arsenin (1977).
Once the $\lambda$-value is attained, the solution $X_\lambda$ of the regularized problem by Tikhonov method is given by:
\begin{equation}
X_\lambda =(A^{T} A+\lambda^2 I)^{-1} A^{T} Y.
\label{solTik}
\end{equation}

In this article we use the Tikhonov regularization method using singular value decomposition 
(see appendix A for details) to deconvolve the distribution of the rotational stellar velocities. 

In the data analysed in this article the L-curve method failed, i. e., we do not obtain the ``L" shape in the L-curve plot, but only the horizontal part of it (see details in Appendix B). For this reason we propose the method described below to chose the Tikhonov factor based on the fact that, when $\lambda \to 0$, $X_{\lambda}$ tends to the exact solution $X$, whereby the difference between two regularized solutions tends to $0$. In Monte Carlo runs (sect. \ref{sec:mcsim}) the Tikhonov factor has been calculated with our proposed method (see below). We proved (sect. \ref{sec:mcsim}) empirically that, Tikhonov estimator we obtained, is unbiased and consistent, both desirables properties of any statistical estimator.

We determine the value of Tikhonov factor, $\lambda$, using the following iterative procedure, which turned out to be faster and efficient to obtain the regularization parameter in case of smooth solutions:
\begin{itemize} 
	\item[i)]We start with an initial value of $\lambda$ ($\lambda=\lambda_0$). 
	\item[ii)] In each following iteration we reduce the value of $\lambda$ by a factor $f$, ($\lambda_j=\,f^j \,\lambda_0$), we use typically $f=0.99$. 
	\item[iii)] At iteration step $j$ we calculate the difference between the correspondent regularization solutions:  $\phi = ||X_{\lambda_j}-X_{\lambda_{j-1}}||$.
	\item[iv)] If $\phi$ is small enough, that is, $\phi<\epsilon$, we stop the iterative process and get the value of $\lambda$. Typically a value of $\epsilon= 10^{-7}$ has been used in this procedure.
\end{itemize}
In appendix B, we show the criteria for selecting $\lambda_0$ and  factor $f$.

\section{Monte Carlo Simulation\label{sec:mcsim}}

In this section we present the results of Monte Carlo numerical simulations, to assess the performance of Tikhonov regularization method when applying to deconvolve rotational velocities distribution from Fredhoml integral. Our Monte Carlo runs consist in $n_{MC}= 1000$ independent replications for each of chosen scenarios, where we considered two specific distributions of rotational velocities. Therefore, we simulate 30 different cases described as follows:

\begin{itemize}
\item[a)]Unimodal Distribution: We choose a Maxwellian distribution
\begin{equation}
f_M(x)= \sqrt{\frac{2}{\pi }}\, \frac{1}{\sigma ^3}\, x^2 e^{-\frac{x^2}{2 \sigma ^2}}, \quad  x>0,
\label{eq1M}
\end{equation}
with parameter $\sigma=8$, which is the same distribution used in Cur\'e et al. ($2014$). Furthermore, we consider 
three different cases, each one including an additive error from a uniform distribution $U[-\sigma_{\epsilon}, \sigma_{\epsilon}]$, with PDF given by $f_{U}(x)\,=\,1/(2\sigma_{\epsilon})$ for $-\sigma_{\epsilon}\leq x\leq\sigma_{\epsilon}$. 
The chosen values of $\sigma_{\epsilon}$ are: $\sigma_{\epsilon} = 0.5, 1, 2\,(km/ s)$.

\item[b)] Bimodal Distribution: For a mixed of two Maxwellian distributions
\begin{equation}
f_{2M}(x)= \sqrt{\frac{2}{\pi }}\,\frac{x^2}{A+B}  \left(\frac{A}{\sigma_1^3} e^{-\frac{x^2}{2 \sigma_1^2}} + \frac{B}{\sigma_2^3} e^{-\frac{x^2}{2 \sigma_2^2}} \right) , \quad  x>0,
\label{eq2M}
\end{equation}
 dispersion parameters are: $\sigma_1=5$ and $\sigma_2=15$, and amplitudes: $A=0.3$ and $B=0.7$. We consider the same  additive error cases as the unimodal distribution.
\end{itemize}

Furthermore, for both (uni and bimodal) cases, we consider five sample lengths $n_s$: $n_s=30, 100, 300, 1\,000, 10\,000$.

For each independent Monte Carlo sample we need to simulate two samples, one from the distribution of the rotational velocities (uni or bimodal) and other for the kernel,  $p(y|x)$, representing the distribution of the inclination angles in the Fredholm integral (Eq. \ref{eq-problema}). Then, we multiply each element of the first sample with the correspondent of the second sample and add the error term. This gives the final sample of $vsin i$ of each scenario. 
The following step is to estimate the PDF of the projected rotational velocities with a Kernel Density Estimator (KDE, Silverman 1986). Using a grid of $n_g$ points we discretized the Fredholm integral obtaining the linear system (Eq. \ref{eq-regterm}). 
With this data we calculate the Tikhonov factor $\lambda$ using the procedure described above and obtained the Tikhonov regularization solution, $X_\lambda$, which is the estimated PDF of rotational speeds.

\begin{figure*}[ht]
	\centering
	\includegraphics[scale=0.55]{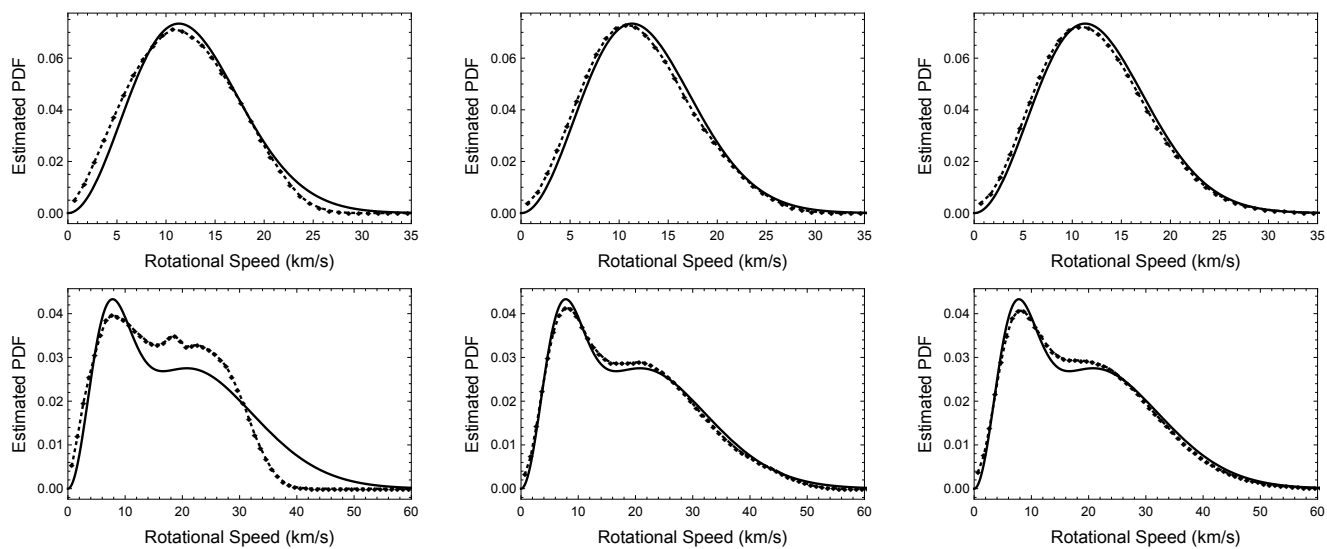}
	\caption{
		Upper panels: Univariate Maxwellian distribution, with parameter $\sigma=8$, is shown in solid line in all upper panels, black squares connected by dashed line represents the mean of the $n_{MC}=1000$ samples of Tikhonov regularization. Results are for:  $n_s=30$ with $\sigma_{\epsilon}=0.5$ (upper left), $n_s=100$ with $\sigma_{\epsilon}=1$ (upper center) and $n_s=1000$ with $\sigma_{\epsilon}=2$ (upper right).
		Lower panels: Bivariate Maxwellian distributions is shown in solid line in all lower panels, with parameters $\sigma_1=5$ and $\sigma_2=15$ and amplitudes $A=0.7$ and $B=0.3$. Black squares connected by dashed line show the estimated PDFs obtained by Tikhonov regularization. Results are for: $n_s=30$ with $\sigma_{\epsilon}=0.5$ (lower left), $n_s=300$ with $\sigma_{\epsilon}=1$ (lower center) and $n_s=1000$ with $\sigma_{\epsilon}=2$ (lower right).}
	\label{Fig1}
\end{figure*}

Figure \ref{Fig1}  upper panels show, in solid line, the original Maxwellian distribution (Eq. \ref{eq1M}) together with the mean estimated PDF of all Monte Carlo simulations (black squares connected by dashed line) for different values of $\sigma_{\epsilon}$ and $n_s$. It is clearly shown that sample lengths of order $n_s \sim 30$ gives acceptable results when compared with the original sample. For larger sample lengths, $n_s \gtrsim 100$, the agreement between original distribution and mean of the estimated PDF is almost exact. Although the mean estimated distribution are slightly shifted to lower velocities. Lower panels of Fig. \ref{Fig1} show the original bimodal mixed Maxwellian distributions (in solid line) together with the mean estimated PDF (black squares connected by dashed line). When a sample length is of order $n_s \sim 30$, a difference between the estimated PDF and the original PDF is observed. Nevertheless, Tikhonov regularized solution retrieves the bimodality and deliver approximately the position of maximum of both components, but gives a wrong estimate of the tail of the original distribution.\\
In the other cases ($n_s \gtrsim 100$) the mean of the Tikhonov regularization solutions is very close to the original mixture of Maxwellian distributions, although the estimated value of the amplitudes is slightly lower (first distribution) and slightly higher (second distribution) than the original one.

In order to quantify the error of the estimated PDF, we calculate (following Cur\'e et al. 2014) the Mean Integrated Square Error (MISE), that is:

\begin{equation}
\mathrm{MISE}=\frac{1}{n_{MC}} \sum_{j=1}^{n_{MC}}\left(\frac{1}{n_{g}}\sum_{i=1}^{n_{g}} (\hat{f_{j}}(x_i)-f(x_i))^2\right).
\end{equation}
where $f(x)$ represent the original distribution function of rotational speeds and $\hat{f}_{j}(x)$ represent the estimated Tikhonov regularization density of the $j$-run in Monte Carlo simulations.

\begin{figure*}[ht]
	\centering
	\includegraphics[scale=0.55]{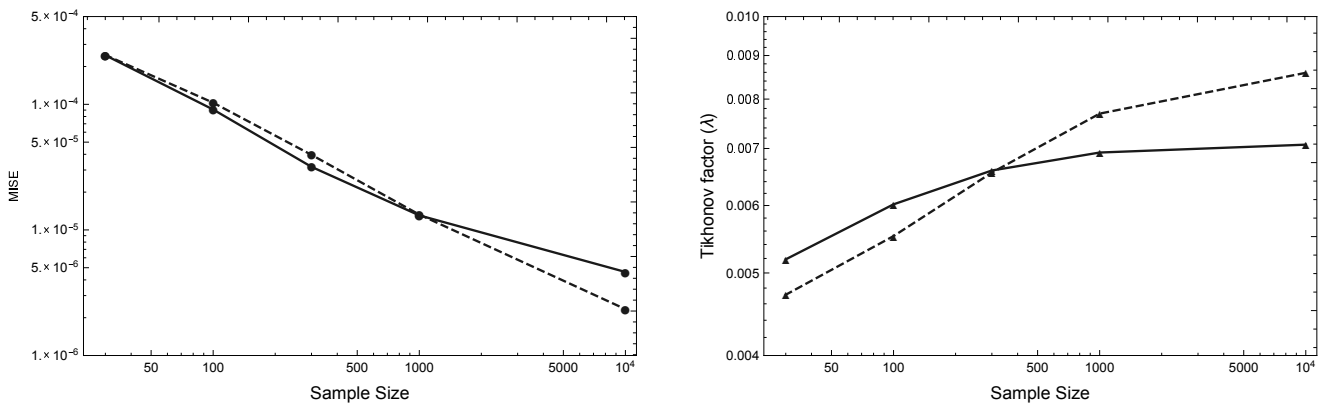}
	\caption{Left panel: The value of MISE (black dots) from the estimated PDF for univariate distributions (solid line) and  bivariate distributions (dashed line) both using  $\sigma_{\epsilon}=0.5$. Right panel show the value of Tikhonov factors $\lambda$ as function of sample size for the cases where $\sigma_{\epsilon}=2$. See text for details.}
	\label{Fig2}
\end{figure*}

In the left panel of Fig. \ref{Fig2}  we plotted the MISE values as function of sample length for $\sigma_{\epsilon}=0.5$.In the other cases ($\sigma_{\epsilon}=1, \, 2$) the MISE value is very similar with values $\mathrm{MISE} \lesssim 10^{-4}$.
Also it can be seen that as increasing the sample size, MISE tends to zero, that is, $\mathrm{MISE}(\hat{f})\tiny{\rightarrow} 0$ when $n_s\tiny{\rightarrow} \infty$. 

The right panel in Fig. \ref{Fig2} shows Tikhonov factors as function of sample size, these factors are of the same order of magnitude for both types of distributions (unimodal and bimodal). Our simulations confirm for all sample lengths and different $\sigma_{\epsilon}$ values that, Tikhonov factor ($\lambda$) is almost independent of the magnitude of the error $\sigma_\epsilon$. Furthermore, since Tikhonov parameter changes slightly as function of sample size,  we can consider the Tikhonov factor is almost independent of the sample length, $n_s$. 

To confirm this result, we have performed MC simulations with a fixed value of $\lambda$. The range of $\lambda$ was from $0.002$ to $0.01$ with a step of $\Delta_ \lambda=0.001$. We calculate the MISE from $n_{MC}=1000$ samples, each with a size of $n_s=1000$ for each value of $\lambda$. For the unimodal
Maxwellian distribution the values of the MISE vary increasing from $7.319 \times 10^{-5}$ to  $7.320 \times 10^{-5}$ for
this range of $\lambda$, a difference almost negligible. In the case of a bimodal Maxwellian distribution the scenario is
very similar using the same range of  $\lambda$, the MISE values vary increasing from $4.717 \times 10^{-5}$ until 
$4.718 \times 10^{-5}$. Similar behaviour is found when $n_s=30, 100, 300, 10000$,  supporting our claim about Tikhonov factor ($\lambda$) is almost independent of the sample size.

By means of the average of the estimated PDFs we can estimate the expected value for the Tikhonov regularization solution. In all cases, the mean of the estimated PDFs is very close to the original unimodal or bimodal distributions, and this mean probability density function is closer to the true PDF when increasing the sample size. 

This fact shows, empirically, that the studied estimator is asymptotically unbiased. Therefore, since MISE tends to zero when $n_s$ tends to infinity, it implies that the variance of the Tikhonov regularization estimator tends to zero as well and hence it is a consistent estimator.

\section{Deconvolving a Real Sample \label{sec:realsample}}
In this section, we perform the following steps: i) Apply Tikhonov regularization method to a sample of measured $v\sin i$ data of cluster stars in order to estimate the rotational velocity probability density distribution, ii) Compare  the application of different methods to deconvolve the velocity distribution together with previous non-parametric results from the literature.

\subsection{Tarantula Sample \label{sec:tarantula}}
We select the Tarantula sample for single O-type stars from the VLT Flames Tarantula Survey, where Ram\'irez-Agudelo et al. (2013) deconvolved the rotational velocity distribution using the Lucy (1974) method (see also Richardson 1972). This sample contains 216 stars with $v \sin i$ data from $40  \, km/s$ up to $610 \, km/s$. Following Ram\'irez-Agudelo et al. (2013), for comparison purposes, we also omitted the two largest values of the sample (outliers). To build the $Y$ vector, we used the KDE method with the following bandwidths (Silverman, 1986, pages 45 and 47):
\begin{eqnarray}
h_1&=&0.79 \, IQR\,  n_s^{-1/5} \\
h_2&=& 0.9 \,\min\{\Sigma,IQR/1.34\} \,n_s^{-1/5},
\end{eqnarray}
here, $IQR$ is the interquartile range and $\Sigma$ is the standard deviation of the sample and $n_s$ is the sample length.\\

\begin{figure*}[ht]
	\centering
	\includegraphics[scale=0.6]{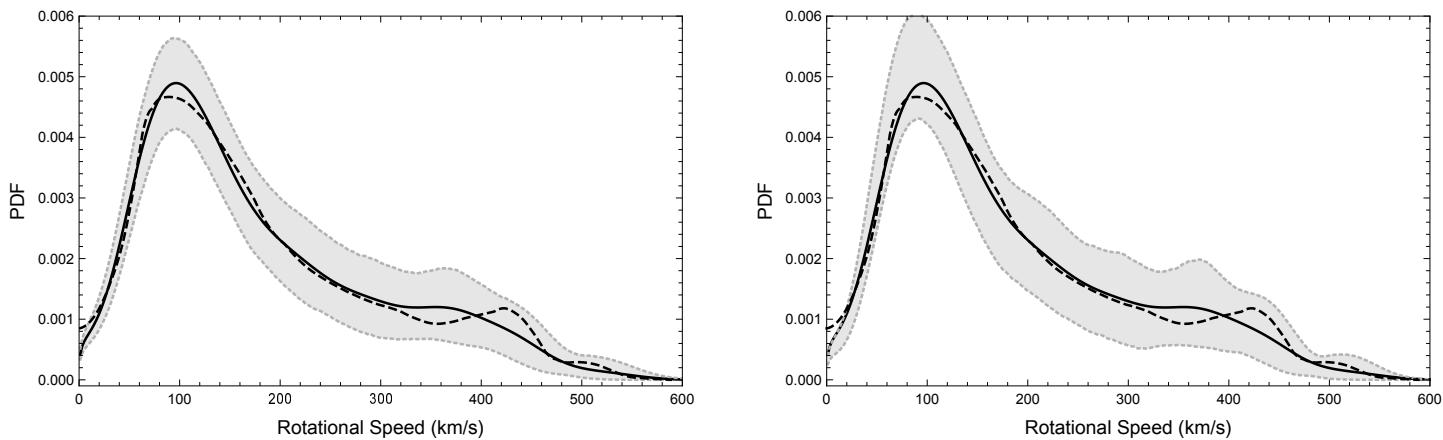}
	\caption{The estimated PDF from Tarantula sample in solid lines. Both panels with $\lambda=0.0174$ and $\Delta x=2\, km/s$, Left panel with bandwidth $h_1=35.676$ and right panel with bandwidth $h_2=30.313$. Gray--shaded regions represent the $2.5\%$ (lower) and $97.5\%$ (upper) confidence intervals calculated by bootstrap method. Dashed lines show the PDF (from Ram\'irez-Agudelo et al. 2013) obtained using Lucy (1974) method.}
	\label{Fig5}
\end{figure*}

Figure \ref{Fig5} shows, in solid line, the rotational velocity distribution after Tikhonov regularization. Our procedure for Tikhonov factor determination gives a value of $\lambda=0.0174$ for a step of $\Delta x= 2\, km/s$. Left panel uses a bandwidth $h_1=35.676$ and right panel a bandwidth $h_2=30.313$. In Fig. \ref{Fig5} we also plotted in light gray the confidence intervals calculated using bootstrap method ($n_{BS}=3000$). The lower is the bandwidth, the wider is the confidence interval. The bump around $ 400 - 450  \, km/s$ is wider in our case ranging from  $\sim 340 \, km/s$ to $  \sim 480  \, km/s$. This discrepancy is probably due to the use of the KDE method with a Gaussian kernel in $Y$.

\subsection{Comparing Results \label{sec:compresults}}
\begin{figure}
	\centering
	\includegraphics[scale=0.38]{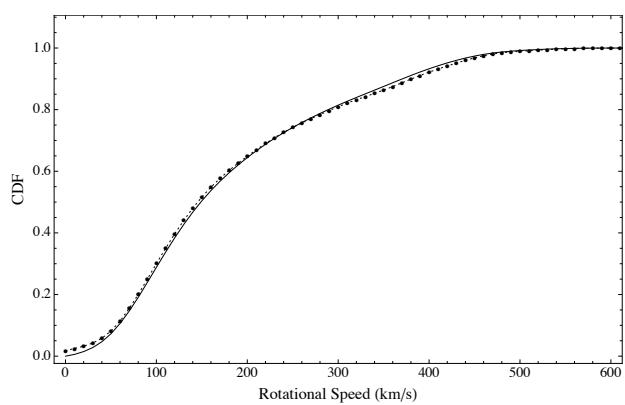}
	\caption{The estimated cumulative rotational velocity distribution function for Tarantula sample (solid line) obtained using Tikhonov regularization using a spacing of $\Delta x=2\, km\, s^{-1}$ for the velocities. Dots connected by dashed line shows the CDF calculated using Cur\'e et al. (2014) method with a spacing of $\Delta x=10\, km\, s^{-1}$.}
	\label{Fig10}
\end{figure}

\begin{figure}
	\centering
	\includegraphics[scale=0.38]{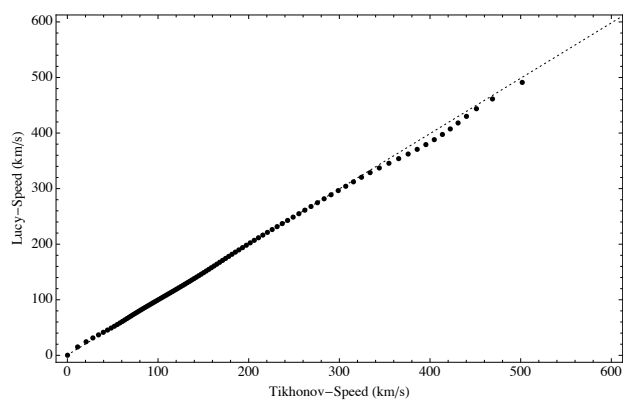}
	\caption{q-q Plot from Tarantula sample, black dots represent the quantiles of each distribution, one calculated using Tikhonov regularization method and the other calculated using the Lucy method (data from Ram\'irez-Agudelo et al., 2013).}
	\label{Fig11}
\end{figure}

\vspace{-.1cm}
For the Tarantula sample, we calculate the CDF by direct integration of the PDF obtained by Tikhonov regularization method  and compare with the CDF calculated by the method described in Cur\'e et al (2014). Figure \ref{Fig10}  shows both CDFs, the agreement between both CDFs is remarkable.In addition to our results for the PDF, Fig. \ref{Fig5} also show in dashed lines the PDF obtained from Ram\'irez-Agudelo et al. (2013, see their Fig. [17]) calculated using Lucy (1974) method. It can be clearly seen that Lucy-PDF lies inside our confidence interval.\\ 
In order to evaluate if both estimated PDFs correspond to the same distribution, we obtained the q--q plot, calculating the respective quantiles. Figure \ref{Fig11} shows the q--q plot of these densities, confirming that both coming from the same probability distribution.

\section{Conclusions}

In this work we have obtained the estimated probability distribution function of 'true' rotational velocities using Tikhonov regularization method.
Furthermore, this estimated PDF uses a Tikhonov parameter $\lambda$ obtained by means of an iterative method with a specific stopping criterion in comparison with the widely used  iterative method of Lucy (1974).\\
Through Monte Carlo numerical simulations we assess the proposed method in two cases: when the rotational velocity distribution is described by a Maxwell distribution and for a mixture of two Maxwell distributions. For each situation different scenarios were evaluated obtaining good results for all of them except for $n_s=30$, when the velocities are described by a mixture of two Maxwellian distributions.\\
This method retrieve the typical rotational velocities distribution for uni- and bimodal distribution. 
We showed, empirically, that the studied estimator is asymptotically unbiased and its variance tends to zero. Furthermore, as measure of goodness of fit, the MISE $\lesssim 10^{-4}$ for all sample sizes and tends to zero when $n_s$ tends to infinity.\\

We apply this method to a set of observed data from Tarantula cluster (Ram\'irez-Agudelo et al. 2013). The estimated PDF from Tikhonov regularization method agreed very well with the PDF obtained using Lucy method, as the q-q plot shows, demonstrating a very good performance to deconvolve rotational velocity distribution (PDF).\\
In comparison with the method that delivers the CDF described in Cure et al. (2014), Tikhonov regularization solution gives, by direct integration of the PDF, almost the same non--parametric estimation of the true underlying cumulative distribution function of  rotational velocities.\\

Summarizing, in  Cur\'e et al. (2014) we developed a method to obtain the CDF of  'true' rotational velocities and in this work we present Tikhonov regularization method to obtain the corresponding PDF directly from Fredhoml integral, both methods calculate in a simple and straightforward way, the PDF or CDF, without any assumptions of the underlying distribution. \\

Future work: We want to develop a general function of the kernel of Fredholm integral, $p(y|x)$, in order to describe an arbitrary orientation of rotational axes. Thus, we can 
study the distribution of rotational speeds relaxing the standard assumption of uniformity of stellar axes.

\begin{acknowledgements}
AC thanks the support from Instituto de Estad\'{i}stica, Pontificia Universidad Cat\'{o}lica de Valpara\'{i}so. PE Thanks the support from Advanced Center for Electrical and Electronic Engineering, AC3E, Basal Fund Conicyt FB0008.
MC thanks the support  Centro de Astrof\'isica de Valpara\'iso and Centro Interdiciplinario de Estudios Atmosf\'ericos y Astroestad\'istica. JC thanks the financial support from project: "Ecuaciones Diferenciales y An\'alisis Num\'erico", Instituto de Ciencias, Instituto de Desarrollo Humano e Instituto de Industria, Universidad Nacional de General Sarmiento. DR acknowledge the support of project PIP11420090100165, CONICET.

\end{acknowledgements}

\begin{appendix}
\section{Tikhonov Regularization Method}

In this Appendix we give a brief description of Tikhonov Regularization Method following closely Burger (2007) and Eggermont (1993).
Suppose that we have a linear system of the form 
\begin{equation}\label{eq:lineal}
A\,X=Y, 
\end{equation}
with a matrix $A\in \mathbb{R}^{n\times n}$, and vectors $X, Y \in \mathbb{R}^n$. Suppose additionally that $A$ is a symmetric positive definite matrix. In this case, from spectral theory for symmetric matrices there exist eigenvalues, $0< \mu_1\leq \cdots \leq \mu_n$ and corresponding eigenvectors $u_i\in \mathbb{R}^n$, with the euclidean norm $||u_i||=1$, such that 
\begin{equation}
A=\sum_{i=1}^{n} \mu_{i} u_iu_{i}^{T},
\end{equation}
\noindent where we consider $u_i \in \mathbb{R}^{n\times1}$. \\
Since the solution of (\ref{eq:lineal}) is given by: 
\begin{equation}
X=\sum_{i=1}^{n} \mu_{i}^{-1} u_iu_{i}^{T}, 
\end{equation}
small eigenvalues of $A$ can cause numerical difficulties when they are arbitrarily close to zero and the problem is ill-posed. The condition number $\kappa:=\mu_n/\mu_1$, is a measure of stability of the system. For simplicity we shall assume that $\mu_n=1$ then $\kappa=1/\mu_1$. When we have data with error $Y_{\delta}$ instead of $Y$, satisfying $||Y_{\delta}-Y||<\delta$, we obtain a solution $X_{\delta}$ and the error in the solution is:
\begin{equation}
||X_{\delta}-X||^2=\sum_{i=1}^{n} \mu_{i}^{-2} |u_{i}^{T}(Y_{\delta}-Y)|^2 \leq \mu_{1}^{-2} ||Y_{\delta}-Y||^2,
\end{equation}
then $||X_{\delta}-X||^2\leq \kappa \delta$.\\
One observes that with increasing condition number the error amplification increases as well. Often the nature of the error is unknown, then it is necessary used a method to solve the linear system that deal with error effects. The regularization methods face this problem efficiently. If matrix $A$ is positive semidefinite, its eigenvalues are non-negative, but it can have a zero eigenvalue. In this case, let $\mu_m$ be the smallest positive eigenvalue, then the solution of (\ref{eq:lineal}) becomes:
\begin{equation}
X=\sum_{i=m}^{n} \mu_{i}^{-1}u_iu_{i}^{T}
\end{equation}
and the problem is solvable if and only if $u_{i}^{T}Y=0$ for $i<m$.\\
For data with error we can use the projection $PY_{\delta}$ onto the range of $A$. This analysis can be extended to general matrix $A\in \mathbb{R}^{n\times m}$ by considering the associated system $A^{T}\,A\,X=A^{T}\,Y$, being that the matrix $A^{T}\,A$ is always symmetric positive semidefinite.

Considering $A$ general, in order to shift away from zero the smallest eigenvalues it seems natural to approximate $A^{T}A$ for a family of matrices $A_\lambda:=A^{T}\,A+\lambda\, I$, whose eigenvalues are $\mu_i + \lambda$, if $\mu_i$ are the eigenvalues of $A^{T}\,A$.\\
We obtain an approximated solution $X_\lambda=A_{\lambda}^{-1}\, A^{T}\,Y$  and for data with error we have $X_{\lambda, \delta}= A_{\lambda}^{-1} \,A^{T} \,Y_{\delta}$. The error of the estimation is then 
\begin{equation}
||X-X_{\lambda, \delta}||^2\leq ||X-X_{\lambda}||^2+||X_{\lambda}-X_{\lambda, \delta}||^2, 
\end{equation}
the first term on the right side corresponds to the approximation error and the second term corresponds to the error in data. Using spectral theory (Burger 2007), we obtain that:
\begin{equation}
||X-X_{\lambda, \delta}||^2\leq \frac{\lambda}{\mu_1 (\mu_1+\lambda)}(||Y_{\delta}||+\delta)+\frac{\delta}{(\mu_1+\lambda)}.
\end{equation}\\
The first term on the right side decreases when $\lambda$ tends to zero while the second term on the right side increases when $\lambda$ tends to zero, thus we have to find an estimation of $\lambda$ that is a compromise between the error of the approximation and the error from measurements.

The solution of the Tikhonov regularization can be obtained also from the Singular Value Descomposition (SVD) of matrix $A$. In this case, we write a general matrix 
$A\in \mathbb{R}^{m\times n}$ with rank $n$ in the form:
\begin{equation}
A^{T} A=\sum_{i=1}^{n} u_{i} \sigma_{i} v_{i}^{T},
\end{equation}
\noindent where $u_{i}$ and $v_{i}$ are orthonormal vector of dimensions $m$ and $n$ respectively, and $\sigma_{i}\geq 0$ are the singular values of $A^{T}A$ such that $\sigma_1\geq \sigma_{2}\geq \cdots \geq \sigma_{n}> 0$.
Under this decomposition the Tikhonov solution is given by:
\begin{equation}
X_\lambda=\sum_{i=1}^{n} f_{i} \frac{u_{i}^{T} Y}{ \sigma_{i}} v_{i},
\end{equation}
where $f_{i}$, $i=1, \cdots, n$ are defined by $f_{i}=\sigma_{i}/(\sigma_{i}+\lambda^{2})$.

As we mentioned in section \ref{sec:method} there are several methods to estimate $\lambda$, the most used are the L-curve Criterion, the Discrepancy Principle and Generalized Cross Validation.

The L-curve is a plot of $\log(||A\,X_{\lambda}-Y||_{2}^{2})$ versus $\log(||X_{\lambda}||_{2}^{2})$, the logarithm of two square euclidean norm, for different values of the Tikhonov factor $\lambda$. This plot has the characteristic $``L"$ shape (see Fig. \ref{Lcurve}). 
According to Hansen (2010) the Tikhonov solution $X_{\lambda}$ can be decomposed as $X_{\lambda}=\bar{X}_{\lambda}+X_{\lambda, e}$, where $\bar{X}_{\lambda}=(A^{T}\,A+\lambda^2 I)^{-1}\, A^{T} Y$ is the regularized version of the exact solution $X$, and $X_{\lambda, e}=(A^{T}\,A+\lambda^2 I)^{-1} \, A^{T} e$ is the solution obtained by applying Tikhonov regularization to the error component $e$.  
For small values of $\lambda$, the error dominates the L-curve because the regularized solution $X_{\lambda}$ is dominated by $X_{\lambda, e}$ and for large values of $\lambda$, $X_{\lambda}$ is dominated by $\bar{X}_{\lambda}$, the unperturbed term. The $\lambda$ chosen is which gets a compromise between the two parts, allocated in the corner of the L-curve.   
The L-curve criterion for choosing the regularization factor is one of the most used methods. The advantages are robustness and ability to manage observations with correlated errors. The limitations of the L-curve are the reconstruction of very smooth exact solutions and to treat with a big amount of data (Hansen 2010).

\section{Determination of Regularization Parameters}
\begin{figure}[ht]
	\centering
	\includegraphics[scale=0.3]{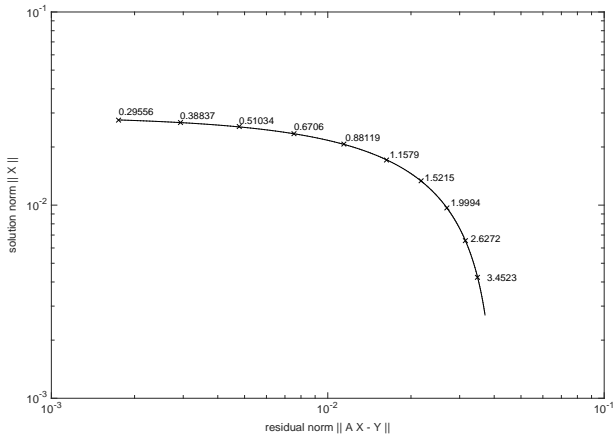}
	\caption{L-curve plot for the data obtained by Monte Carlo sample for a unimodal distribution. Horizontal axis shows  $\log(||A\,X_{\lambda}-Y||_{2}^{2})$, i.e., the residuals of the regularization. Vertical axis shows $\log(||X_{\lambda}||_{2}^{2})$, i.e., the norm of the regularization. Thihkonov parameter  values ($\lambda$) are overplotted to the corresponding data points. It is shown only the horizontal part of the typical 'L' shape, this situation occurs with very smooth exact solution. See text for details. \label{Lcurve}}
\end{figure}

When we apply the L-curve method to different $v \sin i$ samples, the obtained values of the Tikhonov factor ($\lambda$) are 'large'. The reason of these large values is due to the small values of the coefficients in singular value decomposition with almost constant singular values around $1$, having to add to much terms to increase the norm of $X$ (the vertical part of the 'L' shape, see Fig. \ref{Lcurve}). We suspect that the reason of this is  the smoothness of the solution (Hansen 2010). For the Tarantula sample (sect. \ref{sec:realsample}), the Tikhonov parameter delivered by the L-Curve and GCV methods are the same, $\lambda=0.2956$.\\

Here we show how to determine the value of $\lambda_0$ and the choice of factor ($f$) to select the parameters $\lambda$ of the Thikhonov method.

As we stated at the end of section (\ref{sec:method}), we start with a initial value of $\lambda_0$ and calculate the 
Tikonov method to obtain the PDF, ($X_{\lambda}(1)$), then we multiply $\lambda_0$ by a factor $f$ and we obtain a new value of $\lambda=\lambda(1)=\lambda_0 \times f$, and another PDF ($X_{\lambda}(2)$), after applying Tikhonov method. After 'm' iterations we have a set of \{$\lambda(1),\lambda(2),\dots,\lambda(m)$\}.\\
Defining $\phi(j)$ as:
\begin{equation}
\phi(j)=\|X_{\lambda}(j)-X_{\lambda}(j-1)\|
\end{equation}
where $\|\cdot\|$ represent the euclidian norm, after these 'm'iterations we also have a set of \{$\phi(2),\phi(3),\dots,\phi(m)$\}. The iteration stops when the value of $\phi(m)$ is less than certain value $\epsilon$. In our case we choose $\epsilon=10^{-7}$.

\begin{figure}[ht]
	\centering
	\includegraphics[scale=0.75]{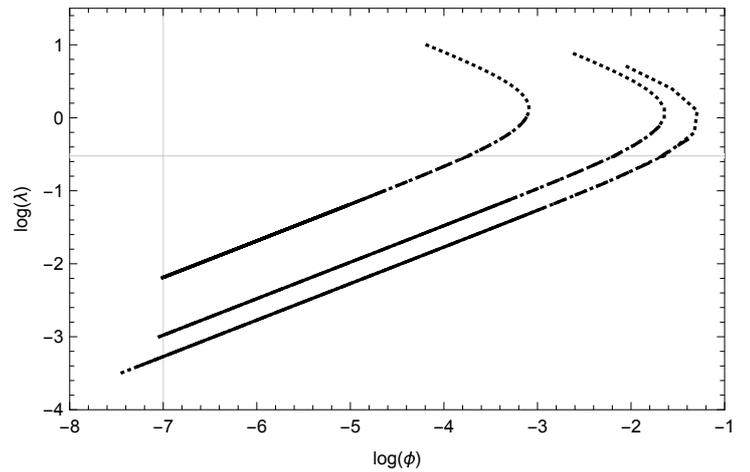}
	\caption{ $\log(\lambda)$ versus $\log(\phi)$. Factor $f$ varies from $f=0.99$ (left), $f=0.75$ (center) to $f=0.5$ (rigth). Each of these curves start with 3 initial values of $\lambda_0$,  dotted line ($\lambda_0=10$), dashed line ($\lambda_0=1$) and solid line ($\lambda_0=0.1$). See text for details.The vertical gray  solid line shows the selected value of $\epsilon=10^{-7}$ as a criterion to finish the iteration process. The horizontal gray solid line shows the value of $\lambda=0.2956$ ($\log(\lambda)=-0.53$) obtained using the L--Curve or GCV methods.
	}
	\label{FigB1}
\end{figure}

Figure \ref{FigB1} shows, $\log(\lambda)$ versus $\log(\phi)$ for different values of $\lambda_0$ and $f$, for the Tarantula sample. The initial values of $\lambda_0$ are: $\lambda_0=10$, shown in dotted line in all 3 curves; $\lambda_0=1$, shown in dashed lines and $\lambda_0=0.1$, in solid lines.\\

For a given value of $f$, all 3 curves are superposed, showing that the final value of $\lambda$ is independent on the starting value $\lambda_0$. Therefore we choose to start our calculations with $\lambda_0=0.1$. On the other hand, the critical parameter here is $f$, the lower is this value, the lower is the final value of $\lambda$, when $\phi(m)\leq \epsilon$. Considering that $\lambda$ is of order $\lambda^2$ in Eq. (\ref{solTik}), a not very small parameter $\lambda$ should be selected in order to have a non-zero regularization term.
Thus we select $f=0.99$ as our default value to obtain the Tikhonov parameter $\lambda$.

It is clearly seen in Fig. \ref{FigB1}, that for $\log(\lambda)=-0.53$, i.e., the value obtained by the L--Curve or GCV method (horizontal gray line), corresponds to a very 'high' value of $\epsilon$. If $f=0.99$, $\epsilon=1.8 \cdot 10^{-4}$, value much larger than $\epsilon=10^{-7}$, which is our criterion to stop this iteration process.

\end{appendix}

\end{document}